\documentclass[10pt,conference]{IEEEtran}
\IEEEoverridecommandlockouts
\usepackage{cite}
\usepackage{amsmath,amssymb,amsfonts}
\usepackage{algorithm}
\usepackage{algpseudocode}
\usepackage{graphicx}
\usepackage{textcomp}
\usepackage{xcolor}
\def\BibTeX{{\rm B\kern-.05em{\sc i\kern-.025em b}\kern-.08em
    T\kern-.1667em\lower.7ex\hbox{E}\kern-.125emX}}

    \IEEEoverridecommandlockouts\IEEEpubid{\makebox[\columnwidth]{ 978-1-6654-3540-6/22~\copyright~2022 IEEE \hfill} \hspace{\columnsep}\makebox[\columnwidth]{ }}
\begin{document}

\title{Power Allocation for Joint Communication and Sensing in Cell-Free Massive MIMO\\

\author{\IEEEauthorblockN{Zinat Behdad, \"Ozlem Tu\u{g}fe Demir, Ki Won Sung, Emil Bj\"ornson, and Cicek Cavdar} 
	{Department of Computer Science, KTH Royal Institute of Technology, Stockholm, Sweden} \\ {Email: \{zinatb, ozlemtd, sungkw, emilbjo, cavdar\}@kth.se
	}
}
\vspace{-1mm}

\thanks{
Results incorporated in this paper received funding from the ECSEL Joint Undertaking (JU) under grant agreement No 876124. The JU receives support from the EU Horizon 2020 research and innovation programme and Vinnova in Sweden.}
\vspace{-4mm}
}
\maketitle

\begin{abstract}
This paper studies a joint communication and sensing (JCAS) system with downlink communication and multi-static sensing for single-target detection in a cloud radio access network architecture. A centralized operation of cell-free massive MIMO is considered for communication and sensing purposes. The JCAS transmit access points (APs) jointly serve the user equipments (UEs) and optionally steer a beam towards the target. A maximum a posteriori ratio test detector is derived to detect the target using signals received at distributed APs. We propose a power allocation algorithm to maximize the sensing signal-to-noise ratio under the condition that a minimal signal-to-interference-plus-noise ratio value for each UE is guaranteed. Numerical results show that, compared to the fully communication-centric power allocation, the detection probability under a certain false alarm probability can be increased significantly by the proposed algorithm for both JCAS setups: i) using additional sensing symbols or ii) using only existing communication symbols.%
\end{abstract}


\section{Introduction}

Joint communication and sensing (JCAS) is envisioned to  become a key technology in future wireless systems by enabling a large number of high-accuracy sensing applications. The main idea of such a system is to share infrastructure, resources, and signals between communications and sensing  to improve the spectral efficiency, sensing performance, and reduce hardware cost \cite{liu2020joint,liu2022survey,zhang2021enabling,wild2021joint}.
JCAS benefits from current developments in the wireless systems, i.e., denser access point (AP) deployment and wider bandwidths \cite{liu2022survey,zhang2021enabling}.

The JCAS research has focused mainly on waveform design and signal processing from the link-level perspective (see  \cite{liu2020joint,buzzi2019using,hua2021transmit,pritzker2022transmit}), in which JCAS is mostly considered with mono-static sensing in a single-cell scenario.
To dispense with the full-duplex capability, bi-static sensing can be considered where the transmitter and receiver are not co-located. Multi-static sensing can also be preferable since it provides a diversity gain due to multiple uncorrelated sensing observations at distributed sensing receivers. To implement multi-static sensing in an existing communication system, cell-free massive MIMO is a promising infrastructure \cite{cell-free-book}, which motivates our work. To fully benefit from cell-free massive MIMO, phase-coherent centralized joint processing is preferred, for which the transmit/receive processing is implemented in a central processing unit. The cloud radio access network (C-RAN) architecture facilitates such centralization and synchronization among the distributed APs \cite{demir2022cell}. The C-RAN architecture has recently gained interest in JCAS systems  \cite{huang2022coordinated}.  

To balance the trade-off between sensing and communication performance due to the shared resources (e.g., time, frequency, power, and space), this work proposes a power allocation algorithm for cell-free JCAS, which maximizes the sensing performance under the condition that the desired communication performance is guaranteed. To the best of our knowledge, this is the first work that considers downlink JCAS in cell-free massive MIMO.
\subsection{Related Work}
  
In \cite{buzzi2020transmit},  mono-static sensing is considered for a single-cell JCAS massive MIMO system. 
A power allocation strategy is proposed to maximize the fairness among the user equipments (UEs) while a minimum signal-to-interference ratio (SIR) is guaranteed for sensing. In contrast, a coordinated power allocation is proposed in \cite{huang2022coordinated}, where a set of distributed transmitter APs serve individually their corresponding UEs in a cellular-type network. In addition, distributed radar sensing is applied to estimate the location of the target by jointly processing the received signals in a central unit.
The proposed power allocation in \cite{huang2022coordinated} aims to minimize the total transmit power while a minimum signal-to-interference-plus-noise ratio (SINR) for each UE and the required Cram\'er-Rao lower bound to estimate the location of the target are guaranteed. 

\subsection{Contributions}

Different from the above-mentioned works, this paper studies a cell-free massive MIMO system, where the UEs are served by all JCAS transmitter APs, and the transmitted signals are simultaneously used for sensing to detect the presence of a target at a certain location. The reflected signals are received at the receiver APs and processed in the edge cloud. Two sensing scenarios are considered: i) only the downlink communication symbols are utilized for sensing;
or ii) a separate sequence of sensing symbols with a certain allocated power are jointly transmitted with the communication symbols. In the latter case, we improve the sensing performance by allocating some resources only for sensing. 
We take a two-step approach to optimize the sensing performance\footnote{The ideal design would be to optimize the transmit power coefficients to  maximize the probability of detection. However, such an approach is rather challenging due to the highly coupled relation between the test statistic and the power coefficients.}. We first maximize the sensing signal-to-noise ratio (SNR), 
under the condition that the target is present, then we use the maximum a posteriori ratio test (MAPRT) detector. It has been discussed in \cite[Chap. 3 and 15]{richards2010principles} that for a given false alarm probability, detection probability increases as SNR increases.
The main contributions of this paper are as follows:
    \vspace{-0.3mm}
\begin{itemize}
\item In the first step, we propose a convex-concave programming-based power allocation algorithm to maximize the sensing SNR under SINR constraints at the UEs.
    \item In the second step, we derive the MAPRT detector to detect a target with known location given knowledge of the optimized power coefficients and only the distribution of the target's radar cross-section (RCS). Our new detector captures all the array response vectors from the transmitter APs to the receiver APs through the target and the respective RCS statistics. The optimal centralized receive processing in the edge cloud is embedded in the proposed detector.
\end{itemize}

\vspace{-2mm}

\section{System Model}
\label{section2}
\begin{figure}[tbp]
\vspace{0.5mm}
\centerline{\includegraphics[trim={0mm 0mm 0mm 0mm},clip,
width=0.35\textwidth]{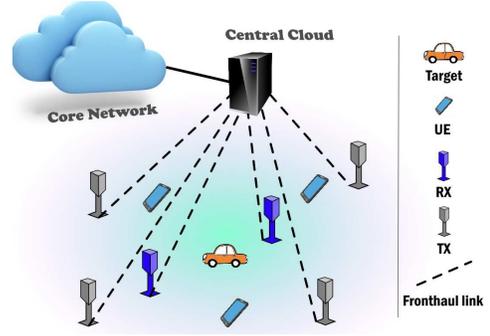}}
\vspace{-3mm}
\caption{Illustration of the JCAS system setup.}
\label{fig1}
\vspace{-5mm}
\end{figure}

We consider a JCAS system with downlink communication and multi-static sensing in a C-RAN architecture, as shown in Fig.~\ref{fig1}. Each AP in the network either serves as a JCAS transmitter or a sensing receiver. The number of transmitter and receiver APs are denoted by $N_{tx}$ and $N_{rx}$ at a given instant, respectively. Each AP is equipped with $M$ antennas that are deployed as a horizontal uniform linear array (ULA). Moreover, all the APs are connected via fronthaul links to the edge cloud and they are fully synchronized. We consider the centralized implementation of cell-free massive MIMO \cite{cell-free-book} for both communication and sensing related processing. The $N_{tx}$ JCAS transmitter APs jointly serve the $N_{ue}$ UEs by transmitting centralized precoded signals. Optionally, they can integrate an additional sequence of sensing symbols that share the same waveform and time-frequency resources with the communication symbols. When this is the case, the transmitter APs jointly steer a centralized beam towards a potential target with a known location. On the other hand, the $N_{rx}$ APs  operate  as sensing receivers by simultaneously sensing the location of the target to determine whether there is a target or not. All processing is done centrally in the edge cloud. 

We assume there is a line-of-sight (LOS) connection between the target location and each AP for simplicity. Assuming the antennas at the APs are half-wavelength-spaced, the antenna array response vector for the azimuth angle  $\varphi$ and elevation angle  $\vartheta$, denoted by $\textbf{a}(\varphi,\vartheta)\in \mathbb{C}^M$, is \cite{bjornson2017massive}
\vspace{-1.5mm}
\begin{equation}\label{a(phi)}
        \textbf{a}(\varphi,\vartheta) =\begin{bmatrix}
          1& e^{j\pi \sin(\varphi)\cos(\vartheta)}& \ldots& e^{j(M-1)\pi\sin(\varphi)\cos(\vartheta)}
        \end{bmatrix} ^T. 
    \end{equation}
 
To focus on the full performance capability of JCAS, we assume  perfect channel state information (CSI) for communication channels as an initial step towards the implementation of JCAS in a C-RAN-assisted cell-free massive MIMO system.\footnote{Channel estimation can be performed  by using either the least-squares (LS) method for distributed radar in \cite{dontamsetti2020distributed} or more advanced channel estimation schemes from the massive MIMO literature.}

 Let $s_i$ denote the zero-mean downlink communication signal for UE $i$ with unit power, i.e., $\mathbb{E}\{|s_i|^2\}=1$. The sensing signal is independent of the UE data signals and is denoted by $s_0$. It has zero mean and $\mathbb{E}\{|s_0|^2\}=1$.
In accordance with \cite{buzzi2019using}, the transmitted signal $\textbf{x}_k[m] \in \mathbb{C}^M$ from transmitter AP $k$ at time instance $m$ can be written as
\vspace{-1mm}
\begin{equation}\label{x_k}
    \textbf{x}_k[m]= \sum_{i=0}^{N_{ue}} \sqrt{\rho_{i}}\textbf{w}_{i,k} s_{i}[m]=\textbf{W}_k \textbf{D}_s[m]\boldsymbol{ \rho}, 
    \vspace{-1mm}
\end{equation}
for $k=1,\ldots,N_{tx}$,
where  $\textbf{w}_{i,k}\in \mathbb{C}^{M}$ and $\textbf{w}_{0,k}\in \mathbb{C}^M$ are the transmit precoding vectors for transmitter AP $k$ corresponding to UE $i$ and the sensing signal, respectively.  
In \eqref{x_k}, $\textbf{W}_k= \begin{bmatrix}
\textbf{w}_{0,k} & \textbf{w}_{1,k} & \ldots & \textbf{w}_{N_{ue},k}
\end{bmatrix}$, ${\textbf{D}}_s[m]=\textrm{diag}\left(s_0[m],s_1[m],\ldots,s_{N_{ue}}[m]\right)$ is the diagonal matrix containing the sensing and communication symbols, and $\boldsymbol{\rho}=[\sqrt{\rho_0} \ \ldots \sqrt{\rho_{N_{ue}}}]^T$.

In line with the C-RAN architecture and the centralized operation of cell-free massive MIMO, the precoding vectors for each UE and the target are jointly selected based on the CSI from all the $N_{tx}M$ distributed transmit antennas. Hence, the precoding vectors for each transmitter AP are extracted from the concatenated centralized precoding vectors given as\vspace{-1mm}
\begin{equation}\label{wi}
        \textbf{w}_i = \begin{bmatrix}
\textbf{w}_{i,1}^T & \textbf{w}_{i,2}^T & \hdots & \textbf{w}_{i,N_{tx}}^T
\end{bmatrix}^T\in \mathbb{C}^{N_{tx}M},
   \vspace{-1mm}
    \end{equation}
    for $i=0,1,\ldots,N_{ue}$.
As can be seen in \eqref{x_k}, there is a common power control coefficient for each UE, $\rho_i\geq 0$, and for the target, $\rho_0\geq0$, since the precoding is a centralized one. The rationale behind this is to preserve the direction of the centralized precoding vectors in \eqref{wi} and, hence, not to destroy its favorable characteristics that is constructed based on the overall channel from $N_{tx}M$ antennas. Using the independence of the data and sensing signals, the average transmit power for transmitter AP $k$ is computed as
\vspace{-1.5mm}
\begin{equation} \label{eq:Pk}
    P_k = \mathbb{E}\{\Vert\textbf{x}_k[m]\Vert^2\} = \sum_{i=0}^{N_{ue}}\rho_i\Vert \textbf{w}_{i,k} \Vert^2, \quad k=1,\ldots,N_{tx}.
    \vspace{-1.5mm}
\end{equation}
Each of the average transmit powers should satisfy the maximum power limit $P_{tx}$, i.e., $P_k\leq P_{tx}$. 

Let $\textbf{h}_{i,k}^*\in \mathbb{C}^M$ denote the channel from transmitter AP $k$ to UE $i$. 
Let us define the sensing/communication channel from all the $N_{tx}M$ transmit antennas to the target and UE $i$ in the network as $\textbf{h}_{0}^*= \begin{bmatrix}
\textbf{a}^T(\varphi_{1},\vartheta_{1})&  \ldots&
\textbf{a}^T(\varphi_{N_{tx}},\vartheta_{N_{tx}})
\end{bmatrix}^T\in \mathbb{C}^{N_{tx}M}$ and  $\textbf{h}_{i}^*=\begin{bmatrix}
\textbf{h}_{i,1}^H& \ldots&
\textbf{h}_{i,N_{tx}}^H
\end{bmatrix}^T\in \mathbb{C}^{N_{tx}M}$, respectively. 
Here, $\varphi_{k}$ and $\vartheta_{k}$ are the azimuth and elevation angles from transmitter AP $k$ to the target location, respectively.

\subsection{Downlink Communication}

The received signal at UE $i$ is given as
\vspace{-1mm}
\begin{align}    \label{y_i}
     y_i[m] =&\sum_{k=1}^{N_{tx}} \textbf{h}^{H}_{i,k}\textbf{x}_{k}[m]+ n_i[m]\nonumber \\
     =&\underbrace{\sqrt{\rho_i}\textbf{h}_{i}^{H}\textbf{w}_{i} s_{i}[m]}_{\textrm{Desired signal}}+ \underbrace{\sum_{j=1,j\neq i}^{N_{ue}}\sqrt{\rho_j}\textbf{h}_{i}^{H}\textbf{w}_{j} s_{j}[m]}_{\textrm{Interference signal due to the other UEs}}\nonumber\\
    &+ \underbrace{\sqrt{\rho_0}\textbf{h}_{i}^{H}\textbf{w}_{0} s_{0}[m]}_{\textrm{Interference signal due to the sensing}}+ \underbrace{n_i[m]}_{\textrm{Noise}} ,
\end{align}
where the thermal noise at the receiver of UE $i$  is represented by $n_i[m] \sim \mathcal{CN}(0,\sigma_n^2)$. The SINR at UE $i$ is given by
\vspace{-1mm}
\begin{align}\label{sinr_i}
     \gamma_i 
     &= \frac{\rho_i\vert \textbf{h}_i^H \textbf{w}_i\vert^2}{\sum_{j=1,j\neq i}^{N_{ue}}\rho_j\vert \textbf{h}_i^H \textbf{w}_j\vert^2+\rho_0\vert \textbf{h}_i^H \textbf{w}_0\vert^2+\sigma_n^2}.
\end{align}

\subsection{Multi-Static Sensing}
\vspace{-1mm}
We consider multi-static sensing, meaning that the sensing transmitters and the receivers are not co-located. Apart from the $N_{tx}$ transmitter APs, the other $N_{rx}$ receiver APs jointly receive the sensing signals for target detection. We assume that the target-free channel between transmitter AP $k$ and receiver AP $r$ is acquired prior to sensing in the absence of the target. Assuming the transmit signal $\textbf{x}_k[m]$ is also known at the edge cloud, the target-free part of the received signal at each receiver AP can be cancelled. After this cancellation, the received signal at AP $r$ in the presence of the target can be expressed as\footnote{In practice, one should also take the cancellation error into account, which is left as future work.}
\vspace{-1mm}
\begin{equation}\label{y_rPrim}
          \textbf{y}_r[m]  = \sum_{k=1}^{N_{tx}} \alpha_{r,k} \underbrace{\textbf{a}(\phi_{r},\theta_{r})\textbf{a}^{T}(\varphi_{k},\vartheta_{k})\textbf{x}_k[m]}_{\triangleq   \textbf{g}_{r,k}[m]}+\textbf{n}_r[m].
          \vspace{-1mm}
\end{equation}
where $\textbf{n}_r[m]\sim \mathcal{CN}(\textbf{0},\sigma_n^2\textbf{I}_M)$ is the receiver noise at the $M$ antennas of receiver AP $r$.
The matrix $\alpha_{r,k} \textbf{a}(\phi_{r},\theta_{r})\textbf{a}^{T}(\varphi_{k},\vartheta_k)$ represents the reflected path through the target where $\phi_r$ and $\theta_r$ are the azimuth and elevation angles from the target location to receiver AP $r$. Here,  $\alpha_{r,k}\sim \mathcal{CN}(0,\,\sigma^{2}_{r,k})$ is the combined sensing channel gain that includes the combined effect of the  path-loss of the path through the target and the RCS of the target\cite{zhao2022joint}. We follow the Swerling-I model for the RCS, in which the velocity of the target is low compared to the total sensing duration. Hence, $\alpha_{r,k}$ is constant throughout the consecutive symbols collected for sensing. Moreover, we assume that they are independent for different APs. In accordance with the previous literature \cite{zhao2022joint}, 
we neglect the paths resulting from multi-reflections from the other objects due to the presence of the target. 

Each receiver AP sends its signal $\textbf{y}_r[m]$, for $r=1,\ldots,N_{rx}$, to the edge cloud. 
    The received signals from all the $N_{rx}$ APs are then concatenated to obtain the overall sensing signal
    \vspace{-1.5mm}
    \begin{equation} \label{y-prime}
  \textbf{y}[m] = \begin{bmatrix}
\textbf{y}_{1}^T[m]&  \ldots&\textbf{y}_{N_{rx}}^T[m]
\end{bmatrix}^T\in \mathbb{C}^{N_{rx}M}.
    \vspace{-1mm}
    \end{equation}
To simplify the analysis, let us first concatenate the unknown sensing channel coefficients for each receiver $r$ as $
    \boldsymbol{\alpha}_r \triangleq [ \alpha_{r,1} \ \ldots \ \alpha_{r,N_{tx}} ]^T \in \mathbb{C}^{N_{tx}}$, then collect all  $\boldsymbol{\alpha}_r$ terms in a vector as
$
\boldsymbol{\alpha}= [ \boldsymbol{\alpha}_1^T \ \ldots \ \boldsymbol{\alpha}_{N_{rx}}^T]^T \in \mathbb{C}^{N_{tx}N_{rx}}$. Moreover, using $\textbf{g}_{r,k}[m]$ from \eqref{y_rPrim}, which is the known part of the received sensing signal, we define the concatenated matrix for receiver AP $r$ and symbol $m$ as $\textbf{G}_r[m] = \begin{bmatrix}\textbf{g}_{r,1}[m]& \ldots &\textbf{g}_{r,N_{tx}}[m] \end{bmatrix}\in \mathbb{C}^{M \times N_{tx}}$.
Then, the overall received signal in \eqref{y-prime} can be expressed as
\vspace{-1mm}
\begin{equation}
    \textbf{y}[m] = \underbrace{\mathrm{blkdiag}\left(\textbf{G}_{1}[m],\ldots,\textbf{G}_{N_{rx}}[m]\right)}_{\triangleq \textbf{G}[m]}\boldsymbol{\alpha}+\textbf{n}[m],
        \vspace{-1.5mm}
\end{equation}
where $\mathrm{blkdiag}(\cdot)$ constructs a block diagonal matrix and $\textbf{n}=\begin{bmatrix}\textbf{n}_1^T[m] & \ldots & \textbf{n}_{N_{rx}}^T[m]\end{bmatrix}^T$ is the concatenated receiver noise.

\vspace{-0.5mm}

\subsection{Transmit Precoding Vectors}
\vspace{-0.5mm}

The unit-norm regularized zero forcing (RZF) precoding vector \cite{cell-free-book}, is  constructed for UE $i$ as $\textbf{w}_{i}=\frac{\bar{\textbf{w}}_{i}}{\left \Vert \bar{\textbf{w}}_{i}\right \Vert}$, where
\vspace{-3mm}
\begin{equation}
     \bar{\textbf{w}}_{i}
     \!=\!\left(\sum\limits_{j=1}^{N_{ue}}\textbf{h}_j\textbf{h}_j^H+\lambda\textbf{I}_{N_{tx}M}\right)^{\!\!-1}\!\!\textbf{h}_i, \ i=1,\ldots,N_{ue},
     \vspace{-2mm}
\end{equation}
and $\lambda$ is the regularization parameter. Note that since the communication symbols also contribute to sensing by the reflected paths towards the target, it is not considered as interference for the sensing target. Hence, we aim at nulling the interference only for the UEs. To null the destructive interference from the sensing signal to the UEs, the sensing precoding vector $\textbf{w}_{0}$ can be selected as the ZF precoder, i.e., by projecting $\textbf{h}_0$ onto the nullspace of the subspace spanned by the UE channel vectors as in \cite{buzzi2019using}.
\vspace{-1mm}
\section{MAPRT Detector for Sensing}\label{section4}
The detection probability, denoted by $P_d$, will be used to measure sensing performance. It is defined as the probability of detecting the target under the condition that the target is present.  In this paper, we use the MAPRT detector based on the already selected power allocation strategy that will be explained in Section \ref{section5}.  We extend the derivation from \cite{guruacharya2020map}, which considers only a single transmitter and single-antenna receivers. Different from the model in \cite{guruacharya2020map}, the transmitted signals are known in our system thanks to the C-RAN architecture. On the other hand, the relation between the received signals at the receiver APs and the unknown RCSs is more complex due to the direction-dependent MIMO channels from the transmitter APs to the receiver APs through the target. Our derived test statistics demonstrate not only how the detector should be implemented but also how the different signals from the receiver APs should be centrally fused in the edge cloud.     
The detection is applied using $\tau$ received symbols. Defining the vectors $\textbf{y}_{\tau}\in \mathbb{C}^{MN_{rx}\tau}$, $\textbf{n}_{\tau}\in \mathbb{C}^{MN_{rx}\tau}$, and $\textbf{g}_{\tau}\in \mathbb{C}^{MN_{rx}\tau}$ constructed by concatenating $\textbf{y}[m]$, $\textbf{n}[m]$, and $\textbf{G}[m]\boldsymbol{\alpha}$, respectively for $\tau$ symbols, the binary hypothesis  used in the MAPRT detector is written as
\vspace{-1.5mm}
\begin{align}\label{hypothesis}
   &\mathcal{H}_0 : \textbf{y}_{\tau}= \textbf{n}_{\tau}, \nonumber\\
  &\mathcal{H}_1 :\textbf{y}_{\tau}=
  \textbf{g}_{\tau}+\textbf{n}_{\tau}.
\end{align}
 The null hypothesis $\mathcal{H}_0$ represents the case that there is no target in the sensing area, 
 while the alternative hypothesis $\mathcal{H}_1$ represents the existence of the target. 
Let $\mathcal{H}\in \{\mathcal{H}_0, \mathcal{H}_1\}$ be the set of hypotheses and $\boldsymbol{\alpha}$ be the vector of unknown channel gains as defined in the system model. The joint RCS estimation and hypothesis testing problem can be written as 
 \vspace{-1mm}
\begin{equation}
    \left(\hat{\boldsymbol{\alpha}}, \hat{\mathcal{H}}\right) = \arg\max_{\boldsymbol{\alpha}, \mathcal{H}} p\left(\boldsymbol{\alpha},\mathcal{H}|\textbf{y}_{\tau}\right),
     \vspace{-1mm}
\end{equation}
where $p\left(\boldsymbol{\alpha},\mathcal{H}|\textbf{y}_{\tau}\right)$ is the joint conditional probability density function of $\boldsymbol{\alpha}$ and $\mathcal{H}$ given the received signal vector $\textbf{y}_{\tau}$. 
Then, the corresponding MAPRT detector can be expressed as 
\begin{align}
    \Lambda = \frac{\max_{\boldsymbol{\alpha}} p\left(\textbf{y}_{\tau}|\boldsymbol{\alpha},\mathcal{H}_1\right) p\left(\boldsymbol{\alpha}|\mathcal{H}_1\right)}{ p\left(\textbf{y}_{\tau}|\mathcal{H}_0\right) }\begin{matrix}
    \mathcal{H}_1\\[-1mm]>\\[-2.5mm] <\\[-1mm] \mathcal{H}_0
    \end{matrix} \lambda_d,
\end{align}
 where $\lambda_d$ is the threshold used by the detector, which is selected to achieve a desired false alarm probability. Following the approach in \cite{guruacharya2020map}, the MAP ratio for our problem is derived as
  \vspace{-1.5mm}
\begin{align}\label{eq_Lambda2}
    \Lambda = C \exp{\left(-\frac{1}{\sigma_n^2}\min_{\boldsymbol\alpha}f(\boldsymbol{\alpha})\right)},
\end{align}
where 
$
    C = \prod_{r=1}^{N_{rx}}\prod_{k=1}^{N_{tx}}\frac{1}{\pi \sigma_{r,k}^2},
$
 \vspace{-1.5mm}
\begin{align}\label{eq:func-alpha}
   f(\boldsymbol{\alpha})=&\boldsymbol{\alpha}^H\left(\sum_{m=1}^{\tau} \textbf{G}^H[m]\textbf{G}[m]+\textbf{D}\right)\boldsymbol{\alpha} \nonumber\\
   &-2\Re\left\{\boldsymbol{\alpha}^H\left(\sum_{m=1}^{\tau} \textbf{G}^H[m] \textbf{y}[m]\right)\right\},
\end{align}
and $\textbf{D}\!=\!\sigma_n^2\textrm{diag}\!\left(\!\sigma_{1,1}^{\!-2},\!\ldots\!,\! \sigma_{1,N_{tx}}^{\!-2},\!\ldots\!,\!\sigma_{N_{rx},1}^{\!-2},\!\ldots\!,\!\sigma_{N_{rx},N_{tx}}^{\!-2}\!\right)$.
Then the channel coefficients $\boldsymbol{\alpha}$ can be estimated as 
\begin{align}\label{eq_alpha}
    \hat{\boldsymbol{\alpha}}\!=\! \left(\sum_{m=1}^{\tau}\textbf{G}^H[m]\textbf{G}[m]+ \textbf{D}\right)^{\!-1}\!\left(\sum_{m=1}^{\tau}\textbf{G}^H[m] \textbf{y}[m]\right).\end{align}
Inserting the optimally estimated $\hat{\boldsymbol{\alpha}}$ into the MAP ratio, we obtain the test statistic $T \triangleq \ln(\Lambda)$ as 
 \vspace{-1.5mm}
\begin{align}
    T &= \ln(C)+\frac{1}{\sigma_n^2}\left(\sum_{m=1}^{\tau}\textbf{y}^H[m]\textbf{G}[m]\right)\nonumber\\
    &\times \left(\sum_{m=1}^{\tau}\textbf{G}^H[m]\textbf{G}[m]+  \textbf{D}\right)^{-1}\!\left(\sum_{m=1}^{\tau}\textbf{G}^H[m]\textbf{y}[m]\right).
    \end{align}
    
Finally, the decision would be
 \vspace{-1mm}
\begin{align}
    \hat{\mathcal{H}}= \left\{\begin{matrix}
\mathcal{H}_1 & \textrm{if}& T\geq \ln(\lambda_d),
\\ 
\mathcal{H}_0 &\textrm{if}& T <\ln(\lambda_d).
\end{matrix}\right.
\end{align}

\section{Power Allocation for JCAS}\label{section5}

We derived the MAPRT detector in Section \ref{section4}. It uses the already selected power allocation strategy, which is the initial stage of our two-stage design approach and elaborated in this section.
We maximize the sensing SNR, denoted by $\gamma_s$, under the condition that the target is present.
The optimization problem can be cast as
\begin{subequations}\label{optimization0}
\begin{align} \label{optimization}
    \underset{\boldsymbol{\rho}\geq \textbf{0}}{\textrm{maximize}} \quad &\gamma_s \\
    \textrm{subject to} \quad 
    &\gamma_i \geq \gamma_c, \quad i=1, \ldots, N_{ue},\label{cona}\\
  & P_k \leq P_{tx},\quad k=1,\ldots ,N_{tx}, \label{conb}
\end{align}
\end{subequations}
where $\gamma_c$ is the minimum required SINR threshold for the UEs and $P_{tx}$ is the maximum transmit power per AP. 
The sensing SNR for the received vector signal in \eqref{hypothesis} is given as
\begin{align}\label{gamma_s}
    \gamma_s = \frac{\mathbb{E}\left\{\Vert \textbf{g}_{\tau}\Vert^2\right\}}{\mathbb{E}\left\{\Vert \textbf{n}_{\tau}\Vert^2\right\}} = \frac{\sum_{m=1}^{\tau}\mathbb{E}\left\{\Vert \textbf{G}[m]\boldsymbol{\alpha}\Vert^2\right\}}{\tau MN_{rx}\sigma_n^2},
\end{align}
where $\mathbb{E}\left\{\Vert \textbf{G}[m]\boldsymbol{\alpha}\Vert^2\right\}$ can be computed as
\vspace{-1.5mm}
\begin{align}
   \mathbb{E}\left\{\Vert \textbf{G}[m]\boldsymbol{\alpha}\Vert^2\right\}  
   &= \sum_{r=1}^{N_{rx}}\sum_{k=1}^{N_{tx}}\mathbb{E}\left\{\textbf{g}_{r,k}^H[m]\alpha_{r,k}^*\alpha_{r,k}\textbf{g}_{r,k}[m]\right\}\nonumber\\
   &\hspace{-16mm}+\sum_{r=1}^{N_{rx}}\sum_{k=1}^{N_{tx}}\sum_{j=1,j \neq k}^{N_{tx}}\mathbb{E}\left\{ \textbf{g}_{r,k}^H[m]\alpha_{r,k}^*\alpha_{r,j}\textbf{g}_{r,j}[m]\right\}.
\end{align}
  Since the edge cloud knows ${\bf x}_k[m]$, we treat $\textbf{g}_{r,k}[m]$ as deterministic. In this case, the last summation in the above equation becomes zero since we assume different RCSs are independent and have zero mean. 

Substituting \eqref{x_k} in the above equation and recalling $\textbf{g}_{r,k}[m]$ from \eqref{y_rPrim}, we can obtain the sensing SNR as 
 \vspace{-1.5mm}
\begin{align}
   \gamma_s=&\boldsymbol{\rho}^T\textbf{A}\boldsymbol{\rho} ,
   \end{align}
    \vspace{-1.5mm}
  where
   \vspace{-1.5mm}
  \begin{align}
 \textbf{A}  =& \frac{1}{\tau MN_{rx}\sigma_n^2}\sum_{m=1}^\tau \textbf{D}_s^H[m] \Bigg(\sum_{r=1}^{N_{rx}}\sum_{k=1}^{N_{tx}}\sigma_{r,k}^2 \textbf{W}_k^H\textbf{a}^{*}(\varphi_{k},\vartheta_{k})\nonumber\\
 &\times\textbf{a}^H(\phi_{r},\theta_{r})\textbf{a}(\phi_{r},\theta_{r})\textbf{a}^{T}(\varphi_{k},\vartheta_{k})  \textbf{W}_k\Bigg)\textbf{D}_s[m].
\end{align}

\vspace{-1mm}
The first set of constraints in \eqref{cona} can be rewritten as  second-order cone (SOC) constraints in terms of $\boldsymbol{\rho}$ as
\begin{align}
&\left \Vert \begin{bmatrix}  a_{i,0}\sqrt{\rho_{0}} & a_{i,1}\sqrt{\rho_{1}} & \!\ldots\! &
\underbrace{0}_{i} & \!\ldots\! & a_{i,N_{ue}}\sqrt{\rho_{N_{ue}}} &
\sigma_n
\end{bmatrix}   \right \Vert \nonumber\\
&\leq \frac{\sqrt{\rho_i} a_{i,i}}{\sqrt{\gamma_c}}, \quad i=1, \ldots, N_{ue}, \label{cona2}
\end{align}
where  
$
 a_{i,j} =
\left \vert \textbf{h}_{i}^H \textbf{w}_{j}\right\vert$ for $i= 1,\ldots,N_{ue}$ and $ j=0,\ldots, N_{ue}
$.
The second set of constraints in \eqref{conb} can also be rewritten in a SOC form from \eqref{eq:Pk} as
\begin{align}
\left \Vert \textbf{F}_k\boldsymbol{\rho}\right \Vert \leq \sqrt{P_{tx}} \label{conb2}
\end{align}
where $\textbf{F}_k =\textrm{diag}\left( \Vert\textbf{w}_{0,k}\Vert, \ldots,\Vert \textbf{w}_{N_{ue},k}\Vert\right)$.
Now, the optimization problem in \eqref{optimization0} can be expressed as a convex-concave programming problem as  
\begin{subequations}\label{optimization2}
\begin{align} 
   &\underset{\boldsymbol{\rho}\geq \textbf{0}}{\textrm{minimize}}  -\boldsymbol{ \rho}^T\Re\{\textbf{A}\}\boldsymbol{ \rho}\\
 &   \textrm{subject to} \quad  \text{\eqref{cona2}, \eqref{conb2}}.
  \end{align}
  \end{subequations}
 Note that since $\textbf{A}$ is a Hermitian symmetric matrix, $\boldsymbol{ \rho}^T\Im\{\textbf{A}\}\boldsymbol{ \rho}$ can be shown to be equal to zero for any real $\boldsymbol{\rho}$. 
 Therefore, \eqref{optimization2} is equivalent to the original problem in \eqref{optimization0}. This problem can be solved using the concave-convex procedure, whose steps are outlined in Algorithm~\ref{alg:fixed-point-uplink}. Under some mild conditions, this algorithm is guaranteed to converge to a stationary point of the problem \cite{lanckriet2009convergence}.
 
\begin{algorithm}
	\caption{Concave-Convex Procedure for Power Allocation} \label{alg:fixed-point-uplink}
	\begin{algorithmic}[1]
		\State {\bf Initialization:} Set an arbitrary initial positive $\boldsymbol{\rho}^{(0)}$ and the solution accuracy $\epsilon>0$. Set the iteration counter to $c=0$.
		 \State $c\leftarrow c+1$.
		\State Set $\boldsymbol{\rho}^{(c)}$ to the solution of the following convex problem (where the previous iterant $\boldsymbol{ \rho}^{(c-1)}$ is taken as constant):
		\begin{subequations}
        \begin{align}
         &\underset{\boldsymbol{\rho}\geq \textbf{0}}{\textrm{minimize}}  -\boldsymbol{ \rho}^T\Re\{\textbf{A}\}\boldsymbol{ \rho}^{(c-1)}\\
         &\textrm{subject to} \quad \text{\eqref{cona2}, \eqref{conb2}}.
        \end{align}
        \end{subequations}
        \State If $\left\Vert \textbf{A}\left(\boldsymbol{\rho}^{(c)}-\boldsymbol{\rho}^{(c-1)}\right)\right\Vert \leq \epsilon $, terminate the iterations. Otherwise return to Step 2.
		\State {\bf Output:} The transmit power coefficients  $\boldsymbol{\rho}^{(c)}$.
	    \end{algorithmic}
\end{algorithm}

\section{Numerical Results}\label{section6}
\begin{figure}[h]
\vspace{3.5mm}
\centerline{\includegraphics[trim={2mm 0mm 9mm 9mm},clip,width=0.4\textwidth]{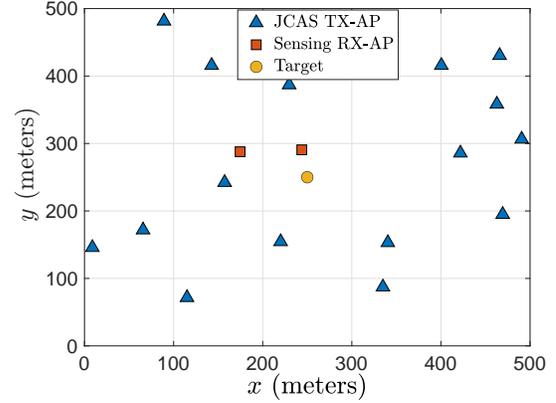}}
\vspace{-3mm}
\caption{The 2D locations of the APs and the target.}
\label{locations} \vspace{-3.5mm}
\end{figure}

\vspace{-2mm}
In this section, numerical results are provided where an area of 500\,m $\times$ 500\,m is considered. The locations of the UEs are randomly generated while the location of the target is fixed and is assumed to be in the center of the area. The locations of the $N_{tx}+N_{rx}$ APs are also uniformly generated, however, it is assumed that they are fixed during the simulation. Among all APs, the $N_{rx}$ closest ones to the target are selected as the sensing receivers. The 2D locations of all the APs and the target are illustrated in Fig.~\ref{locations}. The maximum transmit power per AP is $1$\,W.   
The path loss for the communication channels and sensing channels are modeled by the 3GPP Urban Microcell model, defined in \cite[Table B.1.2.1-1]{3gpp2010further}, and the two-way radar equation, respectively. The carrier frequency, the bandwidth, and the noise variance are set to $1.9$\,GHz, $20$\,MHz, and $-94$\,dBm, respectively. The communication channels follow independently and identically distributed Rayleigh fading and the RCS of the target is modeled by the Swerling-I model where $\alpha_{r,k} \sim \mathcal{CN}(0,\sigma_{rcs}^2)$. 
We set the UE SINR threshold to 20\,dB.  
The detection threshold, i.e., $\lambda_d$ is determined according to the false alarm probability $P_{fa}$ of $0.1$, which is relevant for radar applications \cite{guruacharya2020map}. The numbers of transmitter and receiver APs are $N_{tx}=16$ and $N_{rx}=2$, respectively. The number of symbols that are used for sensing is $\tau=100$ and there are $N_{ue}=8$ UEs, unless otherwise stated.

We compare the sensing performance of i) the  communication-centric design as a baseline algorithm; ii) the proposed JCAS power allocation algorithm ii.a) with and ii.b) without dedicated sensing symbols (``with $s_0$'' and ``w/o $s_0$'' in the figure legends). In the former case, we minimize the power consumption under the same constraints. In the latter case, the sensing SNR is maximized. $P_d$ is computed by averaging 100 random UE locations and channel realizations. For each setup, $1000$ random RCS and $1000\cdot\tau$ random noise realizations are considered for sensing. 

\begin{figure}[h!]
\centerline{\includegraphics[trim={2mm 0mm 9mm 9mm},clip,width=0.4\textwidth]{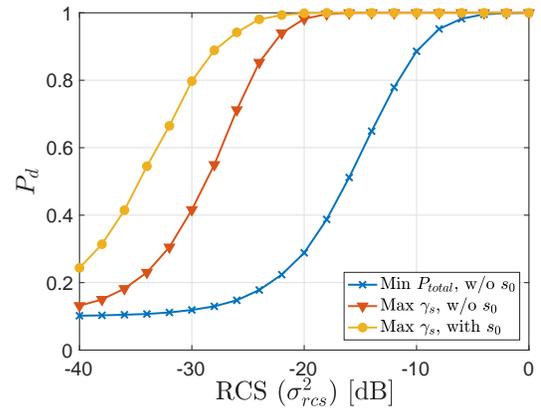}}
\vspace{-3mm}
\caption{$P_d$ vs. RCS for 
$\gamma_c=20$\,dB and $P_{fa}=0.1$.}
\label{PD_RCS} \vspace{-3mm}
\end{figure}
Fig.~\ref{PD_RCS} illustrates the detection probability versus the variance of the RCS.\footnote{Typical values of RCS can span a large range of values starting from $-50$\,dB to $60$\,dB\cite{richards2010principles}. In this paper, we target relatively smaller values of RCS to represent an object relatively hard to detect.} It is shown that, given the same SINR and transmit power constraints, the proposed JCAS algorithm with dedicated sensing symbols can achieve a detection probability of almost $1$ if RCS variance is $\geq-25$\,dB.  At this point, the communication-centric approach results in a detection probability of less than $0.2$ since the sensing requirements are not taken into account. On the other hand, a high sensing performance (e.g., detection probability of $0.8$) is still achievable by maximizing the sensing SNR even without dedicated sensing symbols.  If the RCS is above $-10$\,dB, all the methods achieve the detection probability above $0.9$.

\begin{figure}[t]
\centerline{\includegraphics[trim={2mm 0mm 9mm 9mm},clip,width=0.4\textwidth]{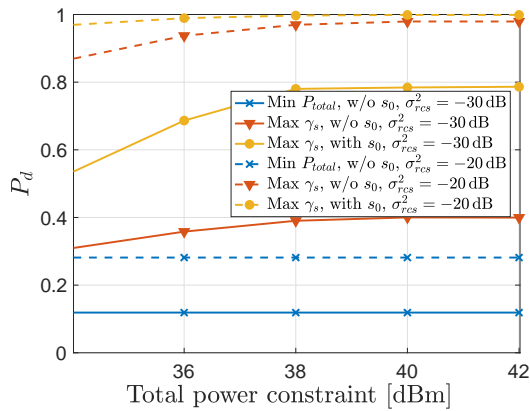}}
\vspace{-3mm}
\caption{$P_d$ vs. the total power constraint for 
$\gamma_c=20$\,dB.} 
\label{PD_PT}
\vspace{-6mm}
\end{figure}

The trade-off between detection probability and the power consumption is depicted in Fig.~\ref{PD_PT}. 
Here, we assume the total power consumption should not exceed a threshold and add the respective constraints to the optimization problems for the three methods. The threshold for this total power constraint is varied along the $x$-axis in the figure. The minimum power threshold is selected a bit higher than the minimum average power obtained by the communication-centric baseline to guarantee feasible power allocation in each random setup.

The performance of the communication-centric design is illustrated by a horizontal line since in this method, the total transmit power is equal to the minimum achievable one following the objective function.
For the proposed JCAS algorithm, the detection probability is growing moderately by sacrificing more power. However, after the power reaches around $38$\,dBm, the curves are quite flat meaning that allowing more power no longer affects the sensing performance. Another interesting result is that the detection probability achieved by the proposed algorithm improves from $0.3$ to $0.9$ compared to the baseline  even when the power threshold is low and no sensing symbols are used.

\begin{figure}[t!]
\centerline{\includegraphics[trim={2mm 0mm 9mm 9mm},clip,width=0.4\textwidth]{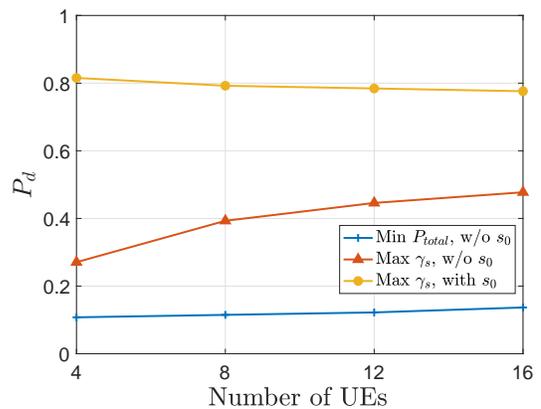}}
\vspace{-3mm}
\caption{$P_d$ vs. the number of UEs for 
$\sigma_{rcs}^2=-30$\,dB and $\gamma_c=20$\,dB.}
\label{PD_UE}
\vspace{-6mm}
\end{figure}

Finally, Fig.~\ref{PD_UE} investigates the impact of the number of UEs on the detection probability. It shows that the more UEs there are in the system, the higher the detection probability becomes when only the communications symbols are utilized for sensing. In contrast, increasing the number of UEs gradually degrades the sensing performance in the case with sensing symbols. 
As the number of UEs increases, less power would be allocated to the sensing symbols to satisfy the increased number of SINR constraints. This restricts the feasible set of the optimization problem, leading to smaller sensing SNR (objective function) under the same per-AP transmit power constraints. As a result, $P_d$ degrades. However, it is partially compensated by increasing the power coming from the communication part. That explains why there is a small decrease in the sensing performance of this algorithm.    

\section{Conclusion} \label{section7}

This paper studied a JCAS system with downlink communication and multi-static sensing in a C-RAN assisted cell-free massive MIMO system. A two-step JCAS approach has been proposed to maximize the sensing performance subject to the UE SINR and per-AP power constraints. 
The numerical results showed that, given a certain false alarm probability, 
maximizing the sensing SNR with/without sensing symbols  would improve detection probability significantly, especially when the RCS is low and/or the number of UEs are relatively high. Dedicated sensing symbols are needed as the RCS goes down, $\leq-20$\,dB.  One may expect that this would occur at the expense of the transmit power. However, we showed that the detection probability increases from $0.3$ to $0.9$ by the proposed JCAS approach compared to the communication-centric approach even when the total power threshold is very low. 
\vspace{-0.5mm}
\vspace{-2mm}

\bibliographystyle{IEEEtran}
\bibliography{IEEEabrv.bib,refs.bib}
\end{document}